\documentclass[aps,prx,twocolumn,superscriptaddress,longbibliography]{revtex4}

\usepackage[english]{babel} 
\usepackage[latin1]{inputenc}
\usepackage[usenames,dvipsnames]{color}
\usepackage{graphicx}
\usepackage{amsmath,amssymb,amsfonts}
\usepackage[colorlinks=true,linkcolor=blue,urlcolor=blue,citecolor=blue]{hyperref}
\usepackage[percent]{overpic}

\begin{document}

\title{Boosting energy transfer between quantum devices through spectrum engineering in the dissipative ultrastrong coupling regime}

\author{Alba Crescente}
\affiliation{Dipartimento di Fisica, Universit\`a di Genova, Via Dodecaneso 33, 16146, Genova, Italy}
\affiliation{CNR-SPIN, Via Dodecaneso 33, 16146, Genova, Italy}
\author{Dario Ferraro}
\affiliation{Dipartimento di Fisica, Universit\`a di Genova, Via Dodecaneso 33, 16146, Genova, Italy}
\affiliation{CNR-SPIN, Via Dodecaneso 33, 16146, Genova, Italy}
\author{Maura Sassetti}
\affiliation{Dipartimento di Fisica, Universit\`a di Genova, Via Dodecaneso 33, 16146, Genova, Italy}
\affiliation{CNR-SPIN, Via Dodecaneso 33, 16146, Genova, Italy}

\begin{abstract}
The coherent energy transfer between two quantum devices (a quantum charger and a quantum battery) mediated by a photonic cavity is investigated, in presence of dissipative environments, with particular focus on the the ultrastrong coupling regime. Here, very short transfer times and high charging power can be achieved in comparison with the usually addressed weak coupling case. Such phenomenology is further magnified by the presence of level crossings appearing in the energy spectrum and which reveal very robust against dissipative environmental effects. Moreover, by carefully control the physical parameters of the model, e.g. the matter-radiation coupling and the frequencies of the system, it is possible to tune these crossings making this device more flexible and experimentally feasible. Finally to broaden our analysis, we assume the possibility of choosing between a Fock and a coherent initial state of the cavity, with the latter showing better energetic performances.
\end{abstract}

\maketitle

\section{Introduction}

The second quantum revolution~\cite{Dowling03} has been one of the most relevant scientific event of the last decades opening the way to the development of quantum technologies. In fact, while the first quantum revolution dealt with the attempt of theoretically explain the fundamental idea of wave-particle duality~\cite{Born26}, the second has taken the new rules of quantum physics and is using them to develop new technologies~\cite{OBrien09, Riedel17, Acin18, Zhang19a, Raymer19, Porta20, Wang20}. Among them, it is worth to mention quantum metrology, quantum communication, quantum computation and, in the last years, also quantum thermodynamics~\cite{Vinjanampathy16, Campisi17, Bera19}. Here, the progressive and increasingly fast miniaturization of the devices, such as quantum thermal machines~\cite{Campisi16, Benenti17, Carrega19, Vischi19}, imposed that the classical laws of thermodynamics could no more be applied and it has been necessary to reconsider them in a regime where quantum effects cannot be neglected. In this context, exploiting the effects of quantum correlations, coherences and entanglement, new devices suitable for energy storage at the quantum level, were considered. In 2013, R. Alicki and M. Fannes introduced for the first time the theoretical concept of quantum battery (QB)~\cite{Alicki13}, a device that allows to store, transfer and release energy with better performances compared to the classical batteries, thanks to collective quantum effects and entanglement~\cite{Campaioli_book, Campaioli17}.
In the last ten years several theoretical works have been devoted to study realistic models and possible experimental implementations, based on simple quantum systems, mostly collections of two-level systems (TLSs), also known as qubits~\cite{Campaioli_book, Andolina18, Farina19}. 
In fact, exploiting two states allows to simply identify the QB as empty when the system is in the ground state and as full when the system is in the excited state. Different scenarios have been considered to charge the QB, i.e. allowing transitions between the empty and full QB. Particular interest has been devoted to classical external fields~\cite{Zhang19, Chen20, Crescente20, Carrega20}, but mostly to quantum chargers, e.g other TLSs~\cite{Andolina18, Le18} or photons trapped into a resonant cavity~\cite{Ferraro18, Andolina19, Crescente20b}. In this direction, possible implementable models have been based on the well known platforms already used for quantum computations, such as artificial atoms~\cite{Le18, Rossini20, Rosa20, Quach20, Santos21, Peng21} and circuit quantum electrodynamics~\cite{Ferraro18, Andolina19b, Dou22}.
First experimental QB works only started to appear in the last two years, the first being the experiment reported in Ref.~\cite{Quach22}, where fluorescent molecules, approximated as TLSs, are placed into a resonant cavity, acting as the quantum charger.
Only later, experimental works based on superconducting qubits~\cite{Hu22} and quantum dots~\cite{Mailette22} have been proposed, increasing even more the interest in the field of QBs.
Moreover, recently implementations in the framework of the IBM quantum machines have been presented, providing another example of functioning QBs~\cite{Gemme22, Gemme23}.

So far, the research on QBs has been mainly focussed on finding efficient ways to store and release energy on demand, used to locally supply it to other miniaturized devices~\cite{Andolina18, Le18, Ferraro18, Centrone23}. At the moment only few works have been devoted to the study of the relevant problem of coherent energy transfer~\cite{Andolina18, Farina19}, and only last year the topic of mediated energy transfer processes has been considered with particular focus on off-resonant conditions, i.e. when the frequencies of each part of the system are not identical~\cite{Crescente22, Crescente23}.
However, these works only considered a weak coupling regime (where the matter-radiation interaction does not exceed $10\%$ of the frequencies of the qubits~\cite{Niemczyk10, Yoshihara16, Kockum19}) between each part of the system, leaving the regime of greater strength matter-radiation couplings ($>10\%$ of the frequencies of the qubits), the so-called strong and ultrastrong coupling (USC) regimes, completely unexplored.
The latter regimes have been investigated in the state transfer literature~\cite{Beaudoin11, Felicetti14, Peng14}, proving that higher couplings could lead to a faster transfer of the quantum state from one qubit to the other. In this direction, great interest has been devoted to the so-called two-qubit Rabi model~\cite{Chilingaryan13, Qu20, Defilippis23}, namely a system where two qubits are coupled to a photonic cavity, but not with each other. In particular, working in the USC regime, despite the additional computational issues related to the failure of the rotating wave approximation (RWA)~\cite{Schweber67, Graham84, Schleich}, the present model shows interesting features such as a sudden population inversion of the photons~\cite{Felicetti14, Peng14} which can lead to a complete and very fast state transfer between the qubits. 
Other than been theoretically interesting, this model has been experimentally implemented in several scenarios, e.g. on resonance with Fock state in the photonic cavity~\cite{Sillanpaa07} and off-resonance with coherent state in the photonic cavity~\cite{HCollard22}. 

Moved by this great interest, in this work we study the mediated energy transfer performances between two quantum devices: a quantum charger and a QB, working in the USC regime, both on and off-resonance. In our description the cavity acts as the mediator of the energy transfer and, to follow the experimental works, both Fock and coherent state will be taken under analysis. We will work in the USC regime, where it is possible to obtain level crossings in the energy spectrum that can be engineered to realize the best working setup. In fact, by changing the initial state of the system it is possible to make relevant one of the different crossings in the energy spectrum for the dynamics of the system. Instead, by tuning the coupling between the different parts of the system or working in the off-resonant condition, it is possible to shift the crossings of the spectrum towards lower values of the coupling strength, moving it from the far USC regime to a more experimentally feasible strong coupling. In general, with the two-qubit Rabi model, working in the USC regime leads to better performances compared to what can be obtained with a weak coupling. Indeed, the presence of the crossings lead to a sudden jump in the transfer times and consequently of the average charging power, namely the ratio between the transferred energy and the minimal time needed to achieve a complete transfer. 
To make the analysis complete and experimentally relevant, we also consider the effects of two external environments (thermal baths) at the same temperature. In particular, in our analysis dissipation is taken in consideration in the framework of the conventional Caldeira-Legget picture~\cite{Weiss, Caldeira83, Leggett87, Ingold}, where one bath is coupled to the cavity and the other one to the QB, to prove the stability of the model. In fact, in real setups it is not possible to neglect the effects of dissipation, which needs to be taken into account. In this direction, it has also been demonstrated that it does not always have detrimental effects on the energy transfer performances of devices~\cite{Carrega20, Rodriguez20, Dias21, Dias23}. Here, we prove that, even in presence of dissipation, where the dynamics is described by the Lindblad formalism~\cite{Lindblad75, Lindblad76}, it is possible to get optimal performances in the USC regime, where the model still present a sudden jump in the charging power. Notice that, the majority of the theoretical works about QB assume unitary dynamics and neglect dissipative effects associated to the coupling with external environments. This approach is usually justified by considering a scale separation between the relevant time evolution of the system and the typical relaxation and dephasing times associated to the coupling with external degrees of freedom~\cite{Devoret13, Wendin17, Crescente20}.
Despite this condition is sometimes fulfilled in simple experimental proposals~\cite{Gemme22, Gemme23}, it is necessary to include dissipative effects to make the description more realistic and experimentally relevant.

The paper is organized as follows. In Sec.~\ref{model} we introduce the model for the cavity mediated energy transfer process, and couplings to the thermal baths, with a particular focus on the initial states. Moreover, the Lindblad formalism is introduced to solve the dynamics in the presence of dissipation. Also, the relevant figures of merit for the energy transfer process are introduced. Sec.~\ref{results} is devoted to the analysis of the results obtained for the closed system dynamics. For the system, we consider the energy spectrum to show its crossings in the USC regime. In addition, we analyze both Fock and coherent initial states, spanning coupling constant from the weak to the USC regimes. At the end of the Section, a possible engineered scheme of the energy spectrum is presented in the off-resonant case. In Sec.~\ref{Dissipation} the stability to dissipation of the previous results is demonstrated for an initial coherent state. Sec.~\ref{conclusion} is devoted to conclusions. Finally App.~\ref{AppA} shows the stability results in the presence of dissipation for the Fock state.


\section{Model}\label{model}

\begin{figure}[h!]
\centering
\includegraphics[scale=0.35]{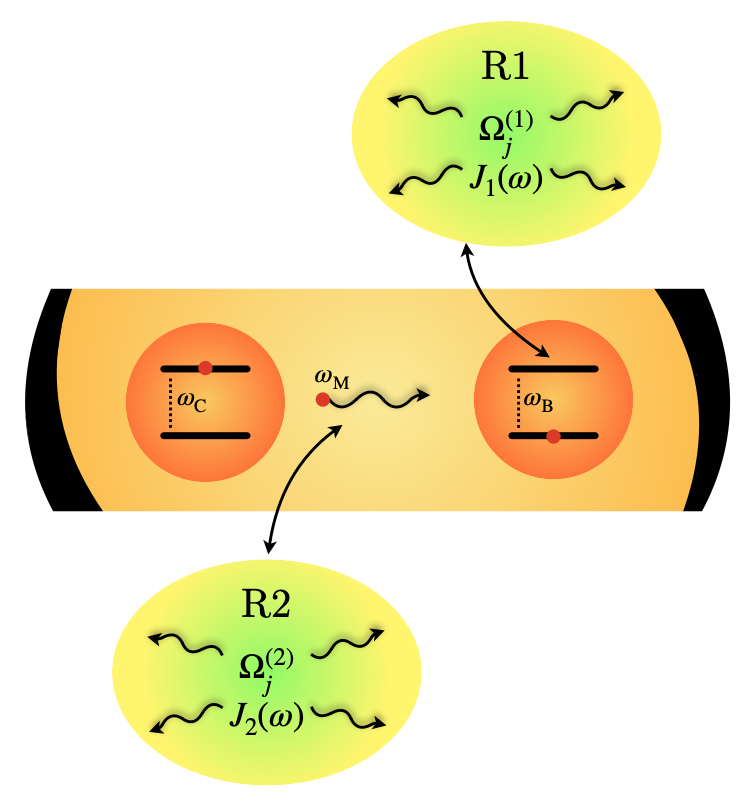} 
   \caption {Scheme of the two-qubit Rabi model in presence of two dissipative baths. Here, two TLSs with level spacings $\omega_{\rm C}$ and $\omega_{\rm B}$ are coupled to a photonic cavity with frequency $\omega_{\rm M}$. The baths are modeled as collections of harmonic oscillators with frequencies $\Omega_j^{(1)}$ and $\Omega_j^{(2)}$ respectively. The first reservoir is coupled to the QB, while the second to the photonic cavity. Both baths are assumed to have Ohmic spectral densities $J_1(\omega)$ and $J_2(\omega)$ [see Eq.~(\ref{ohmic})].}
   \label{fig1}
\end{figure}

In this work, we analyze the energy transfer between two TLSs in terms of the two-qubit Rabi model~\cite{Chilingaryan13, Qu20}, where the first qubit, the quantum charger (C), and the second one, the QB (B) are coupled by means of the photons in the cavity, which play the role of a mediator (M) of the energy transfer (see Fig.~\ref{fig1}). We also consider dissipation in the framework of the conventional Caldeira-Legget picture~\cite{Weiss, Caldeira83, Leggett87, Ingold}, by coupling the QB and the photons in the cavity with two different reservoirs (thermal baths) at the same temperature, modeled as ensembles of harmonic oscillators.
The total Hamiltonian can be written as
\begin{equation}\label{Htot} 
H_{\rm tot}(t)= H(t)+H_{\rm R1}+H_{\rm R2}+H_{\rm RI1}+H_{\rm RI2}. 
\end{equation}

Here, the first term $H(t)$ represents the Hamiltonian of the closed system, composed by the quantum charger, the QB and the cavity. In particular, assuming the conventional dipole interaction between the qubit and the cavity radiation~\cite{Schleich}, it reads (hereafter we set $\hbar=1$)
\begin{equation}\label{Hsys}
H(t)=\frac{\omega_{\rm C}}{2}\sigma_z^{\rm C}+\frac{\omega_{\rm B}}{2}\sigma_z^{\rm B}+\omega_{\rm M}a^\dagger a+ gf(t)(a^\dagger+a)(\sigma_x^{\rm C}+\sigma_x^{\rm B}),
\end{equation}
where $\omega_{\rm C,B}$ are the energy gaps between the ground $|0_{\rm C, B}\rangle$ and the excited states $|1_{\rm C, B}\rangle$ of the two qubits and $\sigma_{x,z}^{\rm C,B}$ are the Pauli matrices along the $\hat x, \hat z$ directions referred to the quantum charger and QB Hilbert spaces respectively. Moreover, $\omega_{\rm M}$ is the frequency of the photons inside the cavity and $a$ ($a^\dagger$) is the annihilation (creation) operator of the photons.
The quantum charger/QB and the photons in the cavity are coupled by means of a coupling strength $g$ with an interaction modulated in time by the switch on and off function 
\begin{equation}
f(t)=\theta(t)-\theta(t-\tau).
\label{ft}
\end{equation}
Here, $\theta(t)$ is the Heaviside step function and $\tau$ is the time interval for which the coupling is turned on. This kind of dynamics can be realized for example by introducing additional elements in the circuit which play the role of quantum couplers~\cite{Sete21, Campbell23, Heunisch23}.

The baths Hamiltonians are written in terms of bosonic creation (annihilation) operators $b_j^{\dagger(i)}$ ($b_j^{(i)}$) as 
\begin{equation} 
H_{{\rm R}i}=\sum_j \Omega_j^{(i)} b_j^{\dagger(i)} b_j^{(i)},\end{equation}
where $\Omega_j^{(i)}$ are the harmonic oscillators frequencies and $i=1,2$ indicates the two different baths.
Notice that, we can neglect dissipative effects on the quantum charger dynamics since the energy transfer process to the QB happens in a very short time. Conversely, it is very important to understand how the loss of photons in the cavity influences the energy transfer performances and to characterize the stability of the energy storing in the QB in presence of dissipation, once the matter-radiation coupling is switched off. 
To do so we consider the interaction Hamiltonians which couple the first bath to the QB and the second bath to the photons, namely 
\begin{eqnarray}
H_{\rm RI1}&=& \sigma_x^{\rm B}\sum_j \lambda_j (b_j^{\dagger(1)}+b_j^{(1)}) \nonumber \\
H_{\rm RI2}&=& (a^\dagger+a)\sum_j \kappa_j (b_j^{\dagger(2)}+b_j^{(2)}).
\end{eqnarray}
The spectral properties of these thermal baths are characterized by the spectral functions~\cite{Weiss}
\begin{eqnarray}
J_1({\omega})&=&\sum_j \lambda_j^2\delta(\omega-\Omega_j^{(1)}) \nonumber \\
J_2(\omega)&=& \sum_j \kappa_j^2\delta(\omega-\Omega_j^{(2)}).
\end{eqnarray}
These equations can be written in the continuum limit and, assuming Ohmic dissipation, they become~\cite{Weiss, Leggett87, Ingold} ($i=1,2$) 
\begin{eqnarray}
J_i({\omega})&=&\alpha_i\omega e^{-\frac{\omega}{\omega_{\rm cut}}}. \label{ohmic}
\end{eqnarray}
Here, $\alpha_1$ and $\alpha_2$ are dimensionless parameters that quantify the dissipation strength and $\omega_{\rm cut}$ is the high frequency cut-off of the baths~\cite{Weiss, Sassetti90, Grifoni96}, which for simplicity is assumed identical for both and considered as the greater energy scale present in the model. 

We now comment on the initial state of the system-bath configuration. Firstly, we assume that, at time $t=0$, the system and the baths are decoupled and  described by the factorized total density matrix
\begin{equation}
\rho_{\rm tot} (0) =\rho(0)\otimes \rho_{\rm R1}(0)\otimes \rho_{\rm R2}(0).
\end{equation} 
As demonstrated in Refs.~\cite{Andolina18, Delmonte21}, the choice of the initial state of the system can have a great impact on the performances of QBs. Therefore, it is important to properly address this point also for more general energy transfer devices. Within this paper, the initial states of the qubits, at $t=0$, will be
\begin{equation} 
|\psi_{\rm C}(0)\rangle=|1_{\rm C}\rangle \quad \quad |\psi_{\rm B}(0)\rangle=|0_{\rm B}\rangle
\end{equation}
for all the considered configurations. This corresponds to the reasonable assumption of a completely full quantum charger and a completely empty QB at the beginning of the energy transfer process.
Different is the situation for the cavity. Here, most of the experimental works in literature have studied a coherent state as initial condition for the photons~\cite{Maring17, Kurpiers18, HCollard22}. Conversely, great part of the theoretical papers addressing QB based on matter-radiation coupling have considered a Fock state~\cite{Ferraro18, Andolina19, Crescente22}. In the present work, for sake of generality, we take into consideration both cases, following Refs.~\cite{Sillanpaa07} and~\cite{HCollard22}. In particular, the initial state of the cavity state is assumed to be
\begin{equation}
|\psi_{\rm M}(0)\rangle=\sum_n \alpha_n|n\rangle,
\end{equation}
where $|n\rangle$ represents a state with $n$ photons and $\alpha_n$ are the associated probability amplitudes. A Fock state with exactly $N$ photons and a coherent state with an averaged number $\bar{N}$ of photons are then characterized respectively by
\begin{equation}
\alpha_n^{\rm F}=\delta_{n,N} \quad\quad\quad
\alpha_n^{\rm C}=e^{-\frac{\bar{N}}{2}}\frac{\bar{N}^{\frac{n}{2}}}{\sqrt{n!}}.
\end{equation}
Summarizing, the initial state of the system can be written as
\begin{equation}\label{initial} |\psi(0)\rangle=|1_{\rm C},0_{\rm B}\rangle\otimes |\psi_{\rm M}(0)\rangle
\end{equation}
with density matrix 
\begin{equation}\label{rho0} \rho(0)=|\psi(0)\rangle\langle \psi(0)|. \end{equation}

Moreover, the reservoirs are at thermal equilibrium with density matrices given by
\begin{equation} \rho_{\rm Ri}(0)=\frac{e^{-\beta H_{\rm Ri}}}{{\rm{Tr}}\{e^{-\beta H_{\rm Ri}}\}}, \end{equation}
with $\beta=1/(k_BT)$ the inverse temperature.

To solve the complete dynamics associated to the Hamiltonian in Eq.~(\ref{Htot}) we will apply the routinely used Lindblad equation~\cite{Lindblad75, Lindblad76}, which to be valid implies weak coupling between system and reservoirs, i.e. $\alpha_{1,2}\lesssim 0.1$, and a Markov approximation~\cite{Petruccione}. This means that the characteristic times associated to the dynamics of the reservoirs $\tau_{\rm R1}, \tau_{\rm R2}$ must be much shorter with respect to the one the system $\tau_{\rm S}$, such that $\tau_{\rm R1}, \tau_{\rm R2}\ll \tau_{\rm S}$.
Under the above conditions it is possible to derive the time evolution of the reduced density matrix $\rho(t)\equiv{\rm Tr}_{\rm R}\{\rho_{\rm tot}(t)\}$ (R stands for the trace over the reservoirs). We have~\cite{Lindblad75, Lindblad76}
\begin{eqnarray}\label{lindblad}
\frac{d}{dt}\rho(t)&=& -i[H(t), \rho(t)] \nonumber \\
&+& \frac{1}{2}\sum_{j=1,2} \bigg[2C_j\rho(t)C_j^\dagger-\rho(t)C_j^\dagger C_j-C_j^\dagger C_j\rho(t)\bigg]. \nonumber \\
\end{eqnarray}
Here, $C_j=\sqrt{\gamma_j} A_j$ ($j=1,2$) are the so-called collapse operators with $A_1=\sigma_x^{\rm B}$ and $A_{2}=a^\dagger+a$, written in terms of the QB and cavity decay rates with
\begin{equation}\label{decr} \gamma_{1,2}=\frac{\pi\alpha_{1,2}\omega_{\rm B,M}^2}{\sqrt{g^2+\omega_{\rm B,M}^2}}\coth\bigg(\frac{\beta\sqrt{g^2+\omega_{\rm B,M}^2}}{2}\bigg)
\end{equation}
which are proportional to $\alpha_1$ and $\alpha_2$ introduced in Eq.~(\ref{ohmic})~\cite{Makhlin03}.

Notice that we have used the numerical tool of the PYTHON toolbox QuTiP~\cite{Johansson13} to solve the dynamics of the system.

Before concluding this Section, we briefly recall the definitions of the quantities of interest to evaluate the energy transfer performances of the device. The energy transferred from the quantum charger to the QB can be written as
\begin{equation}
E_{\rm B}(t)\equiv{\rm{Tr}}_{\rm S}\{\rho(t)H_{\rm B}\}-{\rm{Tr}}_{\rm S}\{\rho(0)H_{\rm B}\}, 
\label{Et} 
\end{equation}
where 
\begin{equation}
H_{\rm B}=\frac{\omega_{\rm B}}{2}\sigma_z^{\rm B}   
\end{equation} 
is the QB Hamiltonian, S stands for the trace over the system, $\rho(0)$ the initial density matrix of the system in Eq.~(\ref{rho0}) and $\rho(t)$ is its time evolved according to Eq.~(\ref{lindblad}). 
Since in realistic situations it is important to transfer as much energy as possible from the quantum charger to the QB in the shortest time, it is also useful to define
\begin{equation} 
E_{\rm B, max}\equiv E_{\rm B}(t_{\rm B, max}), \end{equation}
which corresponds to the maximum of the stored energy in the QB, occurring at the transfer time $t_{\rm B,max}$.
It is also interesting to characterize how much power can be obtained from the QB. In this direction, another two relevant figures of merit are the charging power and the corresponding power evaluated at the maximum transferred energy, defined as 
\begin{equation} 
P_{\rm B}(t)\equiv\frac{E_{\rm B}(t)}{t} \quad\quad\quad P_{\rm B, max}\equiv \frac{E_{\rm B, max}}{t_{\rm B, max}}. 
\label{Pt}
\end{equation}

\section{Closed system results in the USC regime}\label{results}

In the following we present the main results, starting by considering the case where no dissipation is present and $\alpha_1=\alpha_2=0$. The stability of the presented results in presence of dissipative effects will be discussed in the next Section.
Here, we analyze the results concerning the advantages of entering into the USC regime, i.e. $g\approx \omega_{\rm B}$, in order to improve the performances of energy transfer devices. Notice that, previous works have addressed the weak coupling regime~\cite{Crescente22}, where the rotating-wave approximation allows to neglect the counter-rotating terms of the Hamiltonian in Eq.~(\ref{Htot}), i.e. the terms of the form $a\sigma_-$ and $a^\dagger\sigma_+$~\cite{note1}.
Despite very convenient from the computational point of view, this approximation limits the coupling strength to an upper bound of $g\lesssim 0.1\omega_{\rm B}$, leading to a constraint on the energy transfer times~\cite{Crescente22}. To improve the present knowledge about coherent energy transfer processes and widen the perspectives in terms of future applications, we will consider a broader range of coupling spanning from the weak coupling up to the USC, namely $0\leq g\leq0.5\omega_{\rm B}$. The upper bound on the coupling strength is dictated by the Lindblad approximation, which start to fail for $g\gtrsim 0.5\omega_{\rm B}$~\cite{Beaudoin11, Stramacchia19}.
For sake of clarity, the results will be reported in the resonant regime $\omega_{\rm C}=\omega_{\rm M}=\omega_{\rm B}$. This configuration is characterized by the better performances in terms of energy transfer and can be realized experimentally~\cite{Sillanpaa07}. However, at the end of the Section a comment on how to engineer the spectrum of the system by working in the off-resonant regime will be given.

\begin{figure}[h!]
\centering
\includegraphics[scale=0.5]{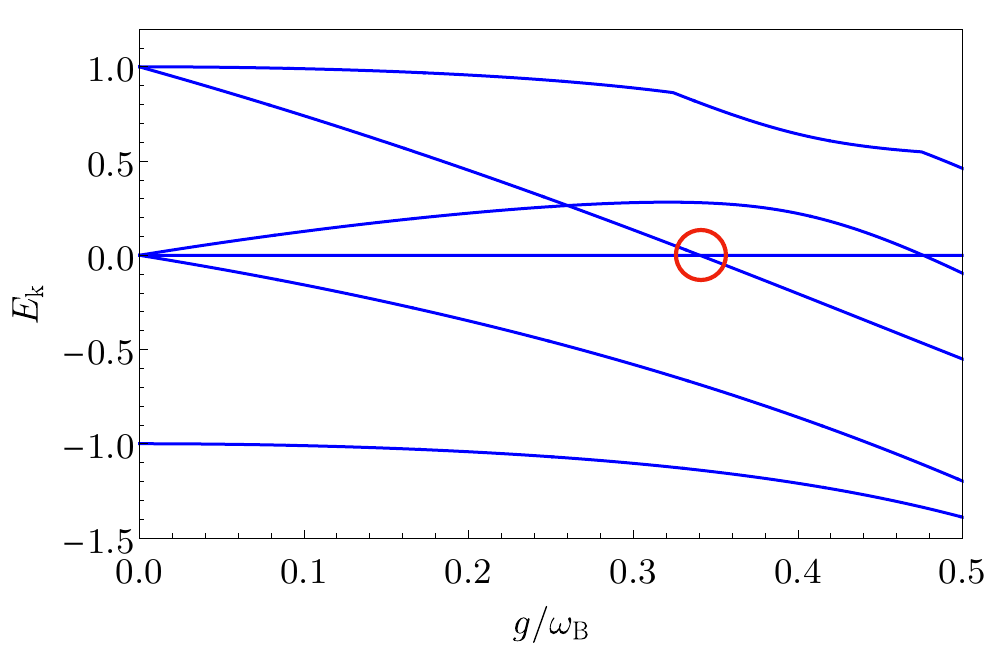} 
   \caption {Eigenvalues $E_k$ of the Hamiltonian in Eq.~(\ref{Hsys}) (in units of $\omega_{\rm B}$) for $f(t)=1$ as function of the coupling constant $g/\omega_{\rm B}$. The model shows level crossings in the USC regime. The crossing at the red circle will be discussed later (see Sec.~\ref{results}). For sake of clarity we have reported only the first six eigenvalues. Other parameters are $\omega_{\rm C}=\omega_{\rm M}=\omega_{\rm B}$, $N_{\rm max} =10 N$.}
   \label{fig2}
\end{figure}

Before analyzing the other figures of merit, it is convenient to study the spectrum of the system of the Hamiltonian in Eq.~(\ref{Hsys}). 
Notice that, the dimension of the Hilbert space of the system grows with the number of photons initially present into the cavity. Therefore, despite some helpful constraints imposed by conservation laws~\cite{Ferraro18, Kirton19}, the Hamiltonian in Eq.~(\ref{Hsys}) cannot be diagonalized analytically and consequently the eigenvalues and eigenstates need to be evaluated numerically. These numerical calculations have been performed by means of the PYTHON toolbox QuTiP~\cite{Johansson13}. In order to constrain the dimension of the Hilbert space, without affecting the reliability of the results, we need to carefully fix a cut-off number of the photons considered in the dynamics $N_{\rm max} =10 N$, with $N$ the number of photons in the cavity. Within this framework it is possible to obtain the energy spectrum in Fig.~\ref{fig2}. 
We notice that, as $g$ increases entering the USC regime, one observes some level crossing~\cite{Peng14}. In the present work we want to show that it is possible to engineer such crossings in order to obtain better energy transfer performances. In fact, by varying the frequencies of the different parts of the system or by considering different couplings between the cavity and the quantum charger/QB it is possible to shift the crossings at smaller value of $g$. 
Moreover by changing the initial state of the system, it is possible to make relevant one of the different crossings in the energy spectrum for the dynamics of the system. 

\begin{widetext}

\begin{figure}[h!]
\centering
\includegraphics[scale=0.46]{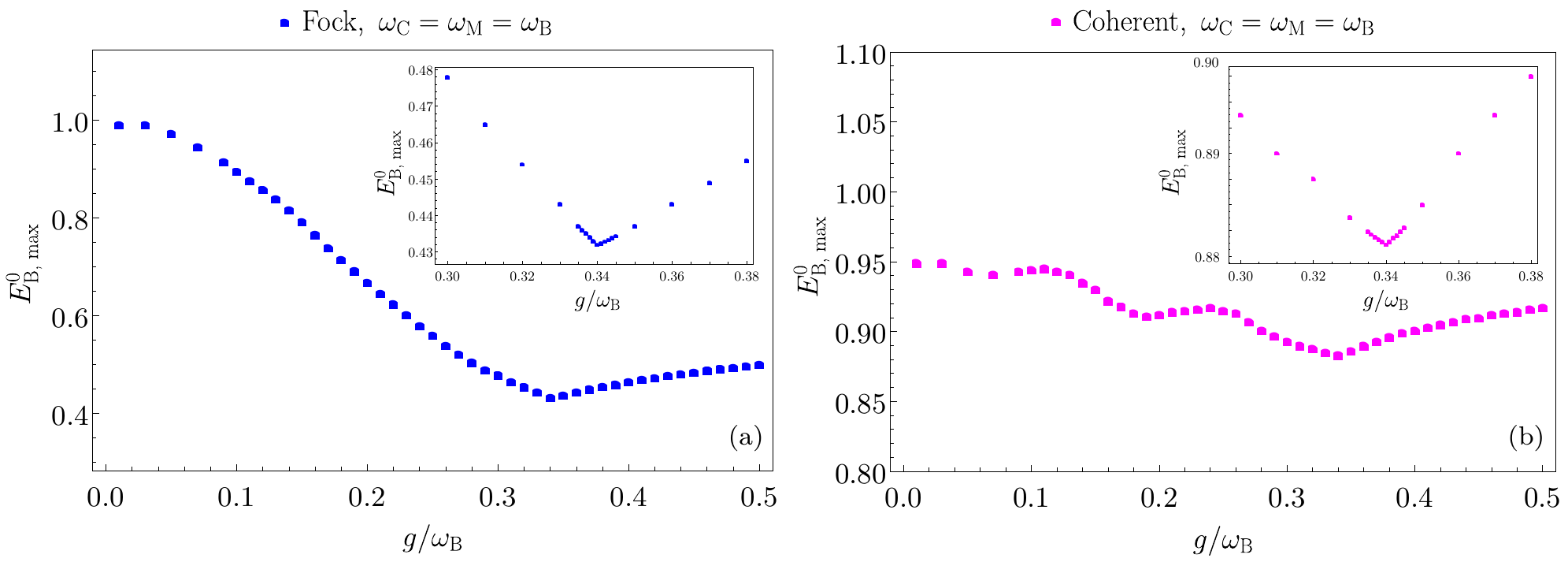} 
   \caption {Behaviour of $E_{\rm B, max}^0$ (in units of $\omega_{\rm B}$) as function of $g/\omega_{\rm B}$ for a Fock state with $N=8$ (a) and a coherent state with averaged number of photons $\bar{N}=8$. Insets show zooms near the value $g^*=0.34\omega_{\rm B}$, where the data shows a cusp. Other parameters are $\omega_{\rm C}=\omega_{\rm M}=\omega_{\rm B}$, $N_{\rm max} =10 N$, $\tau=t_{\rm B, max}^0$ and $\alpha_1=\alpha_2=0$.}
   \label{fig3}
\end{figure}

\end{widetext}

We now consider the effects of such interesting energy spectrum on the different figures of merit.
The behaviour of the maximum of the energy transferred from the quantum charger to the QB as function of the coupling strength is reported in Fig.~\ref{fig3}, considering both a Fock and a coherent state as initial state of the cavity. As a representative case and to guarantee a fair comparison between the two cases we choose $N=8$ as photon number in the Fock state and $\bar{N}=8$ as average number of photons in the coherent state, however similar qualitative results can be obtained for different values. Notice that throughout this Section we use the apex $0$ to indicate that no dissipation is taken into account. Moreover, if not specified differently, the results shown correspond to the choice $\tau=t_{\rm B, max}^0$ in the switch on and off function $f(t)$ in Eq.~(\ref{ft}). 

As a first remark, in Fig.~\ref{fig3} we notice that in the weak coupling regime the Fock state allows a complete energy transfer [panel (a)]. However, as soon as the coupling is increased, the transferred energy drops, reaching its minimum for $g\approx 0.34\omega_{\rm B}$. By further increasing the coupling one has a partial recovery of the transferred energy. A similar qualitative behaviour is obtained for the coherent state. However, as it can be seen in panel (b), this scenario is more stable to the variation of the coupling constant, with a fluctuation of only few percents in the considered range of interaction. It is important to note that both cases present an abrupt change of behaviour for the value $g\approx0.34\omega_{\rm B}$ (see insets of Fig.~\ref{fig3}). This corresponds to the value at which the eigenvalues of the model present the crossing highlighted in Fig.~\ref{fig2}. 
This can be better understood by writing the time evolved state of the system in terms of the eigenvalues $E_k$ and eigenstates $|\varphi_k\rangle$ of the Hamiltonian in Eq.~(\ref{Hsys}) as follows
\begin{equation} 
|\psi(t)\rangle= \sum_{k} c_k(t) |\varphi_k\rangle= \sum_k c_k(0) e^{-iE_k t} |\varphi_k\rangle, 
\label{tev}
\end{equation}
where we have introduced the probability amplitudes $c_k(0)= \langle \varphi_k|\psi(0)\rangle$ and we have only considered the time interval $0<t<\tau$, where the function $f(t)=1$.
By a careful analysis of its elements, it is indeed possible to determine which eigenvalue gives the dominant contribution to the energy transfer. According to this, one has that the energy levels characterized by the crossing occurring the critical value $g^*=0.34\omega_{\rm B}$ are the most relevant (from now on we are going to use the apex $*$ to indicate this peculiar value of the coupling). The role of this energy level crossing is to realize a sudden population inversion of the photons, with important impact at the level of state transfer. Here, we address for the first time the consequences of this peculiar feature at the level of energy transfer. Notice that, by changing the initial condition of the system ($|\psi(0)\rangle$) the coefficients $c_k(0)$ change. As a consequence, also the weight of the different crossings in the dynamics of the system may vary with a consequent lowering of the critical value of $g$ (not shown). This can be a great incentive in engineering the system to obtain the optimal performances of the energy transfer.

To further deepen our analysis, we now demonstrate that this phenomenology has a relevant impact on the energy transfer times and consequently at the level of the power.

\begin{widetext}

\begin{figure}[h!]
\centering
\includegraphics[scale=0.46]{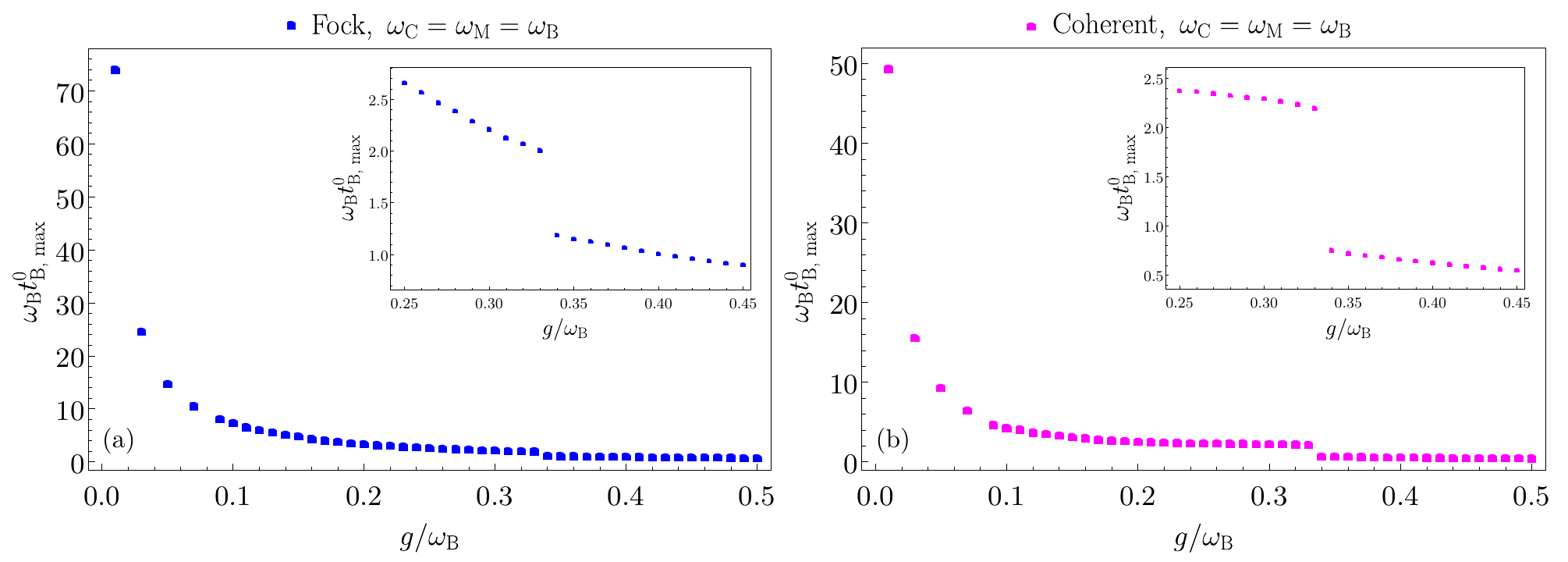} 
   \caption {Behaviour of the energy transfer time  $\omega_{\rm B}t_{\rm B, max}^0$ as function of $g/\omega_{\rm B}$ for the Fock state at $N=8$ (a) and for the coherent state with averaged number of photons $\bar{N}=8$ (b). Insets show zooms near the critical value $g^*=0.34\omega_{\rm B}$, where the transfer time has a jump. Other parameters are $\omega_{\rm C}=\omega_{\rm M}=\omega_{\rm B}$, $N_{\rm max} =10 N$, $\tau=t_{\rm B, max}^0$ and $\alpha_1=\alpha_2=0$.}
   \label{fig4}
\end{figure}

\end{widetext}

The time required to transfer energy from the quantum charger to the QB gets smaller increasing the coupling $g$~\cite{Andolina18, Delmonte21}, as shown in Fig.~\ref{fig4}, and consequently it is very short in the USC regime. Moreover, the energy transfer in the Fock state [panel (a)] is typically considerably slower with respect to the one in the coherent case [panel (b)] .

Apart from these general considerations, it is interesting to look closely at what happens near the critical value $g^*=0.34\omega_{\rm B}$. Here, an abrupt reduction of the energy transfer times occurs for both the Fock and coherent state, as can be seen in the insets of Fig.~\ref{fig4}. It is quite remarkable to see how also this quantity shows a sudden drop in the USC regime, in correspondence of the level crossing, meaning that the proper engineering of the spectrum can also lead to improvements at the level of the transfer times. 
In an energetic perspective, such a sudden decrease of the transfer times can be directly related to an enhancement of the average charging power.

\begin{widetext}

\begin{figure}[h!]
\centering
\includegraphics[scale=0.35]{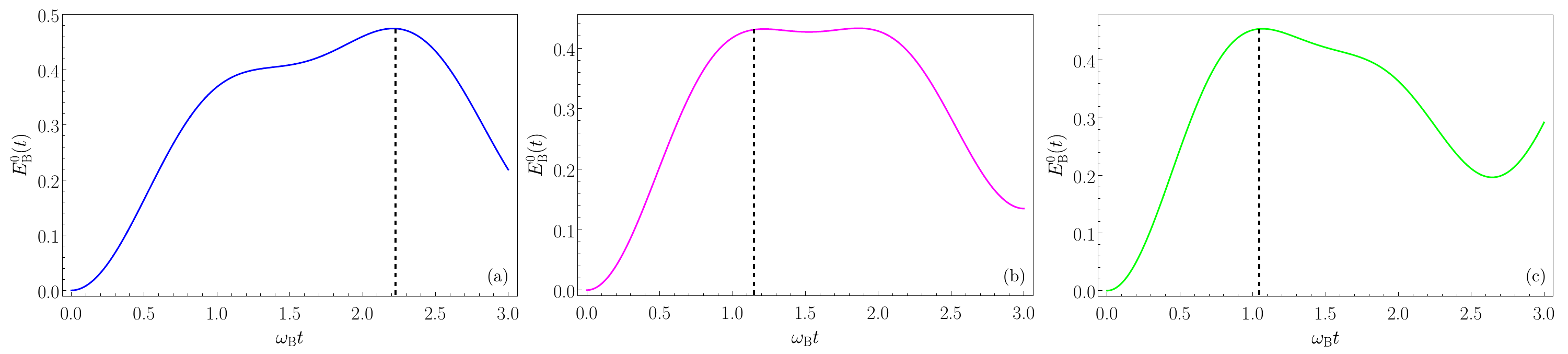} 
   \caption {Behaviour of $E_{\rm B}(t)$ (in units of $\omega_{\rm B}$) as function of $\omega_{\rm B}t$ for a Fock states with $N=8$. We consider three different values of the coupling: $g=0.3\omega_{\rm B}$ (a), $g=g^*=0.34\omega_{\rm B}$ (b) and $g=0.38\omega_{\rm B}$ (c). The black dashed lines represent the positions of the maximum of the transferred energy $t_{\rm B, max}$. Other parameters are $\omega_{\rm C}=\omega_{\rm M}=\omega_{\rm B}$, $N_{\rm max} =10 N$ and $\alpha_1=\alpha_2=0$.}
   \label{fig5}
\end{figure}

\end{widetext}

Before addressing this relevant point, it is useful to better clarify why this reduction occurs. This can be done by analyzing the time evolution of the energy transferred from the quantum charger to the QB, shown in Fig.~\ref{fig5}. We only report the case of the Fock state, but an analogous behaviour is observed also for the coherent state. Notice that, to better enlighten the behaviour of the maximum of the transferred energy, we have considered $\tau\gg t_{\rm B, max}^0$ in Eq.~(\ref{ft}), corresponding to a situation where the matter-radiation coupling is switched on for a long time with respect to the maximum of the transferred energy. With the aim of discussing the behaviour of the transferred energy across $g^*=0.34\omega_{\rm B}$, we compare the cases of coupling constants near $g^*$.
Doing so it is possible to observe that the first two maxima of the transferred energy progressively  exchange their role. In fact, while for $g < g^*$ [panel (a)] the second maximum is the most pronounced, when the critical value is reached the two maxima have exactly the same value [panel (b)]. Finally, for a greater coupling $g>g^*$ [panel (c)],  the first maximum is always the most pronounced and the energy transfer occurs in a shorter time.
This justifies the jump in the transfer time in Fig.~\ref{fig4}, observed at $g=g^{*}$. Notice that, this phenomenology is a peculiarity of the USC regime, further strengthening the interest in exploring this range of parameters.      

\begin{widetext}

\begin{figure}[h!]
\centering
\includegraphics[scale=0.45]{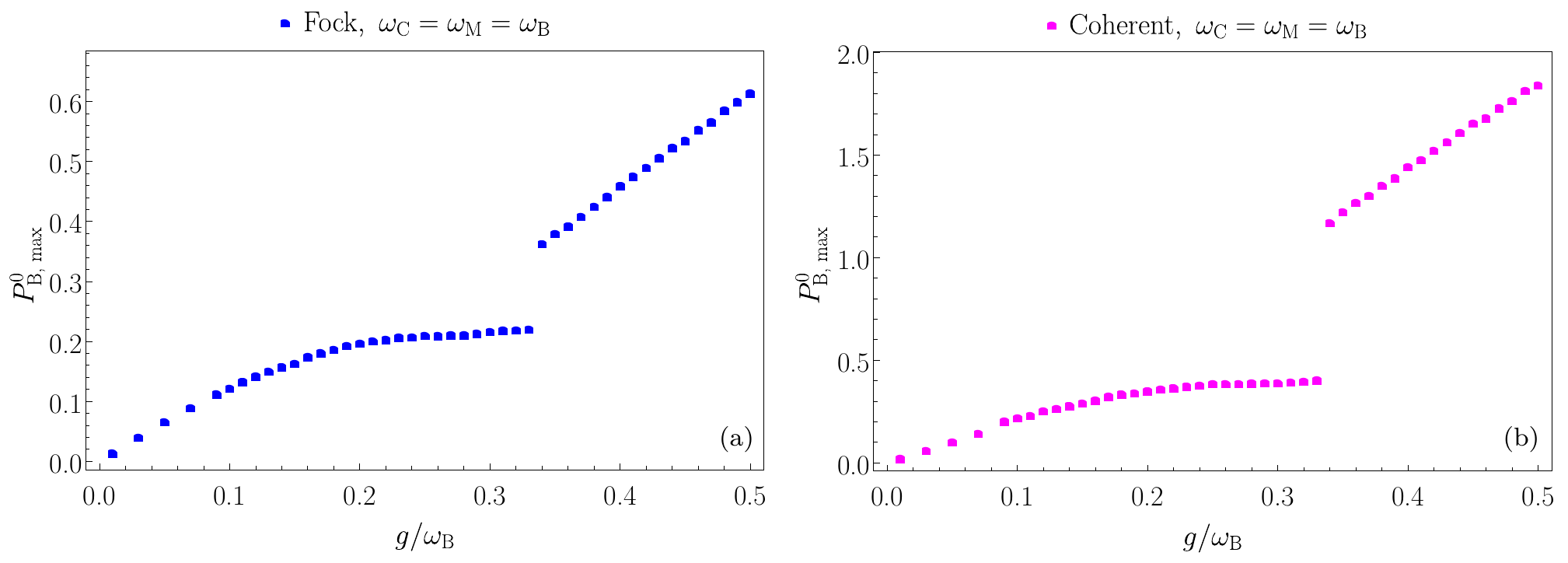} 
   \caption {Behaviour of $P_{\rm B, max}^0$ (in units of $\omega_{\rm B}^2$) as function of $g/\omega_{\rm B}$ for a Fock state at $N=8$ (a) and for a coherent state with average number of photons $\bar{N}=8$ (b). Other parameters are $\omega_{\rm C}=\omega_{\rm M}=\omega_{\rm B}$, $N_{\rm max} =10 N$, $\tau=t_{\rm B, max}^0$ and $\alpha_1=\alpha_2=0$.}
   \label{fig6}
\end{figure}

\end{widetext}

Let's now analyze how the behaviour observed in $t_{\rm B, max}^0$ influences the average charging power. In Fig.~\ref{fig6} we show the evolution of the maximum of the average charging power as a function of the coupling, defined in Eq.~(\ref{Pt}), for both a Fock and a coherent initial state. At $g^*=0.34\omega_{\rm B}$ we observe in both cases a sudden enhancement in the value of $P_{\rm B, max}^0$. 
Due to the previous considerations, one has that the Fock state is less performant compared to the coherent state. Indeed, the former [panel (a)] jumps from $P_{\rm B, max}^0\approx0.22\omega_{\rm B}^2$ to $P_{\rm B, max}^0=0.36\omega_{\rm B}^2$ across the critical coupling while the latter [panel (b)], shows a discontinuity from $P_{\rm B, max}^0\approx0.40\omega_{\rm B}^2$ to $P_{\rm B, max}^0=1.17\omega_{\rm B}^2$. Here, the power increases of almost a factor $3$ by slightly changing the coupling constant. Even more interestingly, the average charging power grows almost linearly in $g$ after the critical value. This leads to a remarkable improvement of the average charging power in the USC regime with respect to what observed at weak coupling couplings.

\begin{widetext}

\begin{figure}[h!]
\centering
\includegraphics[scale=0.5]{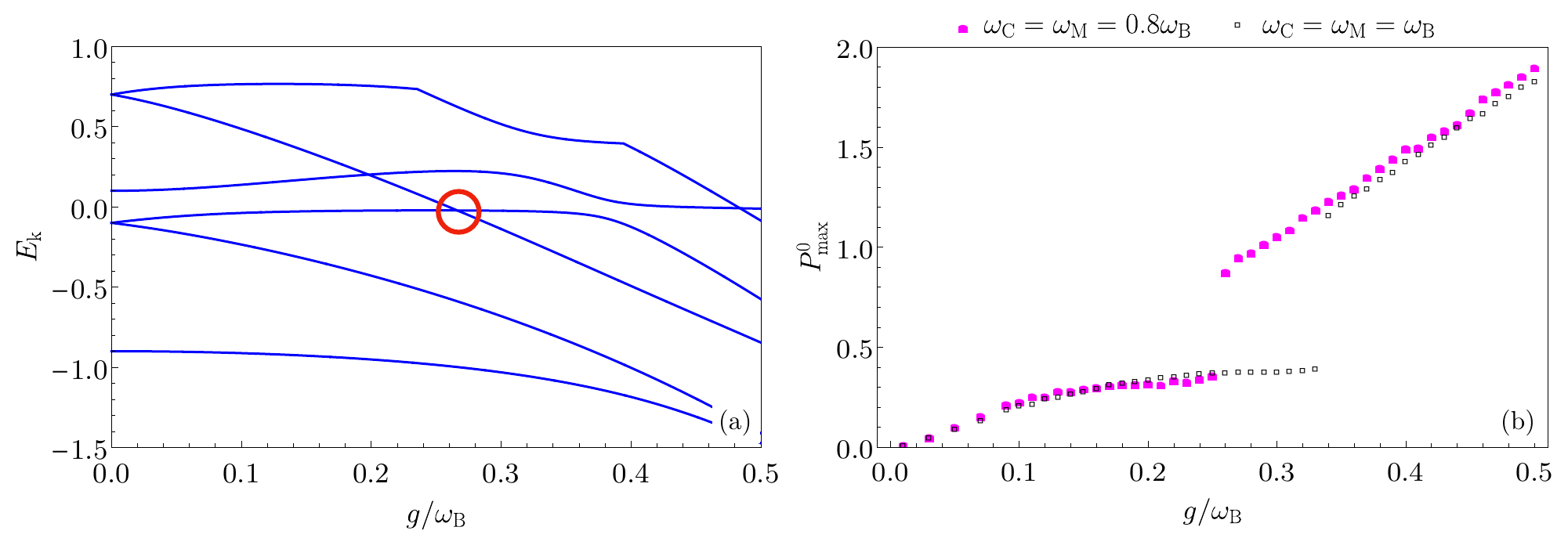} 
   \caption {Panel (a): Eigenvalues $E_k$ of the Hamiltonian in Eq.~(\ref{Hsys}) (in units of $\omega_{\rm B}$) when the system is off-resonance $\omega_{\rm C}=\omega_{\rm M}=0.8\omega_{\rm B}$ in the interval $0< t < \tau$ as function of the coupling constant $g/\omega_{\rm B}$. The red circle indicates the critical coupling $g^*_{off}=0.26\omega_{\rm B}$. For sake of clarity we have reported only the first six eigenvalues. Panel (b): Behaviour of $P_{\rm B, max}^0$ (in units of $\omega_{\rm B}^2$) as function of $g/\omega_{\rm B}$ for a coherent state with average number of photons $\bar{N}=8$ for the off-resonance regime $\omega_{\rm C}=\omega_{\rm M}=0.8\omega_{\rm B}$ (full magenta squares) compared to the resonant case $\omega_{\rm C}=\omega_{\rm M}=\omega_{\rm B}$ (opened black squares). Other parameters are $N_{\rm max} =10 N$, $\tau=t_{\rm B, max}^0$ and $\alpha_1=\alpha_2=0$.}
   \label{fig7}
\end{figure}

\end{widetext}

Before concluding this analysis, it is useful to consider how it is possible to engineer the position of the level crossings, which play a crucial role in the power performance. Among the different possibilities discussed previously, the most relevant from an experimental point of view is to consider an off-resonant regime. This is because it is not always possible to obtain exactly identical qubits or to have photons with the same frequencies as the energy separation of the TLSs. In this direction, we now consider the off-resonant regime $\omega_{\rm C}=\omega_{\rm M}=0.8\omega_{\rm B}$, proposed in the experiment in Ref.~\cite{Sillanpaa07} and also discussed in Ref.~\cite{Crescente22}. Here, it useful to understand how the energy spectrum of the Hamiltonian in Eq.~(\ref{Hsys}) is modified. As we can see from Fig.~\ref{fig7} (a) the eigenvalues of the Hamiltonian are different from the ones in Fig.~\ref{fig2}, with the system on resonance. In particular, the crossings are shifted at lower values of the coupling, where the circled one, representing the critical values for this scenario is obtained for $g^*_{\rm off}=0.26\omega_{\rm B}$ compared to the previous one $g^*=0.34\omega_{\rm B}$. However, to be relevant for an energetic application, it is also necessary to also analyze the charging performances in this regime. We focus on the coherent state since, we have shown that it is more performant compared to the Fock state. Moreover, as the relevant figure of merit we analyze the maximum of the average charging power, reported in Fig.~\ref{fig7} (b) and we give a comparison between the off-resonant and resonant regime. 
Firstly, it is possible to observe that the two scenarios have the same qualitative behaviour, again $P_{\rm B, max}^0$ has a sudden boost at the critical value of the coupling. However, in the off-resonant regime this happens for a lower value of the coupling ($g^*_{\rm off}=0.26\omega_{\rm B}$), meaning that in the interval $0.26\omega_{\rm B}\leq g<0.34\omega_{\rm B}$, this configuration has way better performances compared to the resonant one. For the other values of the coupling the two scenarios have almost the same performances, with the off-resonant regime showing slightly higher values of $P_{\rm B, max}^0$, since the transfer times are in general shorter when the system is off-resonance~\cite{Crescente22}.

From the above results, it is evident that the structure of the eigenvalues of the Hamiltonian in Eq.~(\ref{Hsys}) (especially their crossings) has a great impact on the energy transfer performances of the device. To be flexible and to optimize the performance we have shown that the level crossings can be engineered by changing the parameters of the system. In the considered cases, the crossings at $g^*=0.34\omega_{\rm B}$ and $g^*_{\rm off}=0.26\omega_{\rm B}$ (in the USC regime) lead to a reduction of the energy transfer time and to a consequent enhancement of the average charging power. Moreover, these effects are further enlighten by considering a coherent states as initial state for the photonic cavity.


\section{Stability to dissipative effects}\label{Dissipation}

We now study the stability of the different figures of merit in a more realistic case, in presence of dissipation. For sake of simplicity, we assume the two baths to have the same temperature $\beta \omega_{\rm B}=10$, compatible with experimental values~\cite{Sillanpaa07, HCollard22}. We underline that the temperature of the bath is a relevant parameter that influences the QB and cavity decay rates $\gamma_1$ and $\gamma_2$ in Eq.~(\ref{decr}). In fact, if one chooses a smaller value of $\beta$ (high temperature) the performances of the device are strongly affected by dissipative effects, leading to a very poor energy transfer.

Here, the results will be presented only for the coherent state which, in absence of dissipation, has shown the best performances. The results for the Fock states are commented in Appendix~\ref{AppA}.

\begin{widetext}

\begin{figure}[h!]
\centering
\includegraphics[scale=0.36]{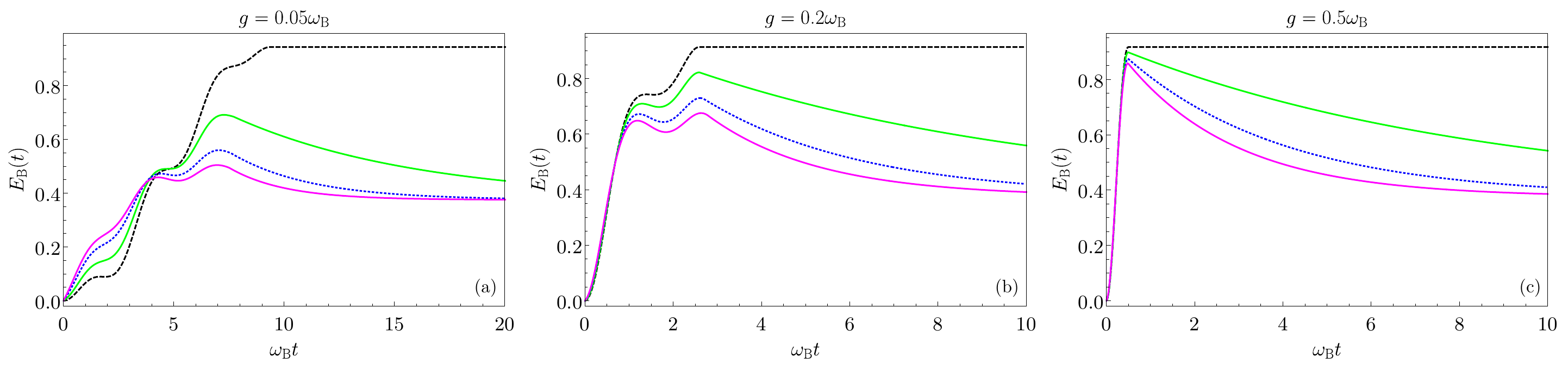} 
   \caption {Behaviour of $E_{\rm B}$ (in units of $\omega_{\rm B}$) as function of $\omega_{\rm B}t$ for a coherent state with average number of photons $\bar N=8$ and three different values of the coupling: $g=0.05\omega_{\rm B}$ (a), $g=0.2\omega_{\rm B}$ (b) and $g=0.5\omega_{\rm B}$ (c). We consider, at fixed $\alpha_2=0.01$, different values of $\alpha_1$: $\alpha_1=0.03$ (full green curves), $\alpha_1=0.07$ (dotted blue curves) and $\alpha_1=0.1$ (full magenta curves). The dashed black curves are the reference case in absence of dissipation ($\alpha_1=\alpha_2=0$). Other parameters are $\omega_{\rm C}=\omega_{\rm M}=\omega_{\rm B}$, $N_{\rm max} =10 N$, $\beta \omega_{\rm B}=10$ and $\omega_{\rm cut}=500\omega_{\rm B}$.}
   \label{fig8}
\end{figure}

\end{widetext}

It is interesting to analyze the stability with respect to dissipation for different values of the couplings ranging from the weak to the USC regime.
In this direction, in Fig.~\ref{fig8} the behaviour of $E_{\rm B}(t)$ is reported for three relevant examples [$g=0.05\omega_{\rm B}$ (a), $g=0.2\omega_{\rm B}$ (b) and $g=0.5\omega_{\rm B}$ (c)]. While the cavity is supposed to have the same dissipation strength $\alpha_2$ throughout the whole analysis, consistently with the possibility to realize very stable cavities~\cite{Sillanpaa07, Scarlino19}, the QB is supposed to be more affected by the action of the environment and consequently we analyze the effects of having different $\alpha_1$. Notice that, the coupling between the baths and the QB and cavity are chosen within the regime of validity of the Lindblad equation, i.e. $0\leq \alpha_{1,2}\lesssim 0.1$~\cite{Petruccione}. Moreover, all the results are compared with the case where no dissipation is present ($\alpha_1=\alpha_2=0$).

First of all, we observe that the energy transfer process is strongly affected by dissipation in the weak coupling regime [panel (a)], even for very small dissipation strength $\alpha_1 =0.03$ and $\alpha_2=0.01$. This gets progressively worse by increasing the coupling $\alpha_1$ between the bath and the QB. However, when the system approaches the USC regime, the dissipative effects become less important. 

In fact, at $g=0.2 \omega_{\rm B}$ [panel (b)], it is possible to obtain up to $\sim80\%$ of the total energy transferred to the QB, at $\alpha_1=0.03$, compared to $\sim90\%$ without dissipation.
Even better is the case where $g=0.5\omega_{\rm B}$ [panel (c)], where dissipative effects have only a marginal impact on the energy transfer process, even considering different dissipative rates for the QB. 
In general, it is also worth to note that choosing different $\alpha_1$ implies that, after reaching the maximum of the transferred energy and switching off the coupling between the parts of the system, the energy approaches the thermal equilibrium value at different times, since with higher $\alpha_1$ comes a faster decay rate. 

This analysis demonstrate that it is better to work in the USC regime, where dissipation only plays a minor role. This is a consequence of the fact that at USC the time scales associated to the energy transfer processes are very short with respect to the dynamics induced by the coupling with the external environment. Moreover, it is obviously convenient to have a low dissipative rate associated to the QB to obtain a more stable storing of the energy into the QB. 

\begin{figure}[h!]
\centering
\includegraphics[scale=0.5]{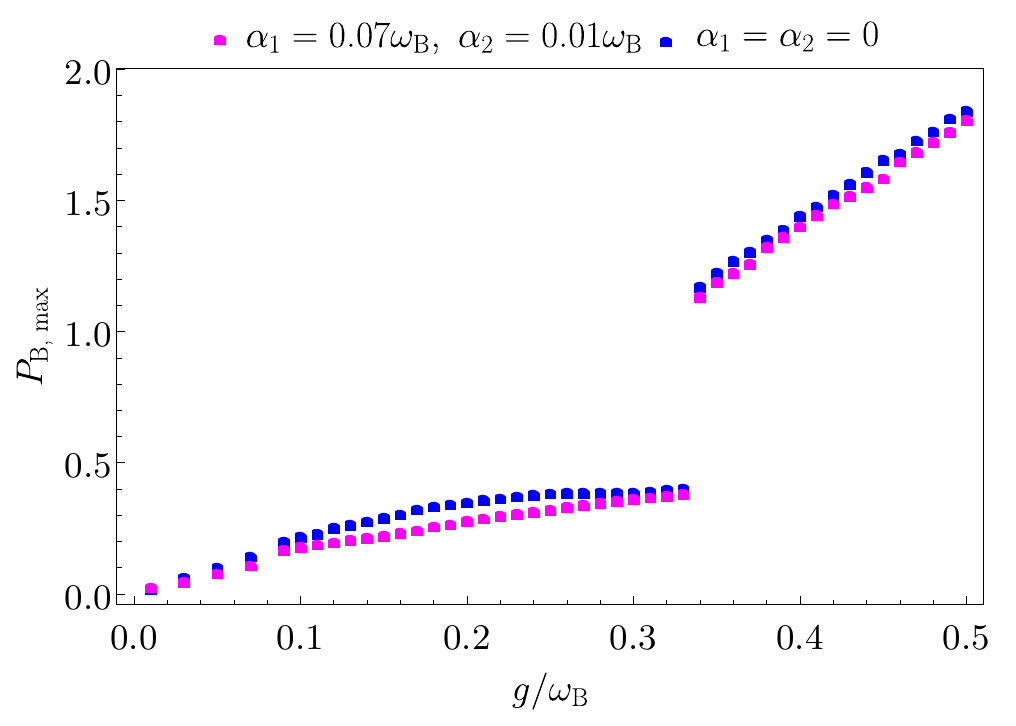} 
   \caption {Behaviour of $P_{\rm B, max}$ (in units of $\omega_{\rm B}^2$) as function of $g/\omega_{\rm B}$ for a coherent state with average number of photons $\bar{N}=8$, for the dissipative case at $\alpha_1=0.07$ and $\alpha_2=0.01$ (magenta square) in comparison with the case with no dissipation at $\alpha_1=\alpha_2=0$ (blue square). Other parameters are $\omega_{\rm C}=\omega_{\rm M}=\omega_{\rm B}$, $N_{\rm max} =10 N$, $\beta \omega_{\rm B}=10$ and $\omega_{\rm cut}=500\omega_{\rm B}$.}
   \label{fig9}
\end{figure}

We now discuss the dissipative effects on the average charging power. In particular, we focus on the representative case with $\alpha_1=0.07$ and $\alpha_2=0.01$ and compare it to the non dissipative case (see Fig.~\ref{fig9}). 

As a general remark it is possible to see that, for these values of the parameters, dissipation has only marginal effects on the value of the $P_{\rm B, max}$. In particular, at weak couplings and in the USC regime the data almost coincides with the non dissipative case, while in the regime between $0.1 \omega_{\rm B}\lesssim g\lesssim 0.25\omega_{\rm B}$ there is a discrepancy between the dissipative and non dissipative case.
Moreover the more relevant feature of the average charging power, namely the jump at $g^*=0.34\omega_{\rm B}$ discussed above, remains unaffected.
It is important to notice that these results are a consequence of the small values of dissipative rates consistent with the Lindblad formalism and motivated by state of the art experiments~\cite{Sillanpaa07, Scarlino19}. However, one would expect that, if higher values of dissipation are considered it should be possible to realize an avoided crossing in the eigenvalues and consequently lose the sudden jump in the power~\cite{Peng14}, leading to a loss in the performances of the device.

In conclusion, the coherent state shows optimal energy transfer performances also in presence of dissipation, particularly in the USC regime, adding another motivation to engineer devices in such conditions by controlling and mitigating dissipative effects.


\section{Conclusion}\label{conclusion}

The present work has been devoted to the analysis of a coherent energy transfer between two quantum devices, namely a quantum charger and a quantum battery, mediated by a photonic cavity. The analysis is brought out in a wide range of coupling strengths, ranging from the weak coupling to the ultrastrong coupling regime. In the latter case the model shows crossing in the energy spectrum, that we demonstrated can be engineered in order to optimize the performance of the energy transfer. Indeed, this peculiar behaviour has a great impact on the different figures of merit. In particular, in the presence of level crossings, the transfer time has a sudden jump at the critical value of the coupling, which also impacts the average charging power, that doubles in the ultrastrong coupling regime compared to the weak coupling one.
Moreover, choosing different initial states has a great impact on the performances. In fact, we have shown that a coherent state for the cavity has better performances compared to the Fock state.
In addition, we have proved the robustness of the model to dissipation. In fact, especially considering a coherent state for the cavity, the presence of two environments, coupled to the cavity and to the quantum battery, does not have strong detrimental impact on the dynamics, mostly in the ultrastrong coupling regime.
This analysis opens the possibility of engineering an energy transfer setup for quantum batteries, where working in the ultrastrong coupling regime allows to obtain better results compared to the conventional ones obtained in the weak coupling scenario. Moreover, being the two-qubit Rabi model experimentally established, this should pave the way for the implementation of our model in the very near future.


\begin{acknowledgments}
Authors would like to acknowledge the
contribution of the European Union-NextGenerationEU through the ``Quantum Busses
for Coherent Energy Transfer'' (QUBERT) project, in the framework of the Curiosity
Driven 2021 initiative of the University of Genova and through the ``Solid State Quantum
Batteries: Characterization and Optimization'' (SoS-QuBa) project, in the framework
of the PRIN 2022 initiative of the Italian Ministry of University (MUR) for the National
Research Program (PNR).
\end{acknowledgments}


\appendix

\section{Dissipative effects on Fock states}\label{AppA}

\begin{widetext}

\begin{figure}[h!]
\centering
\includegraphics[scale=0.36]{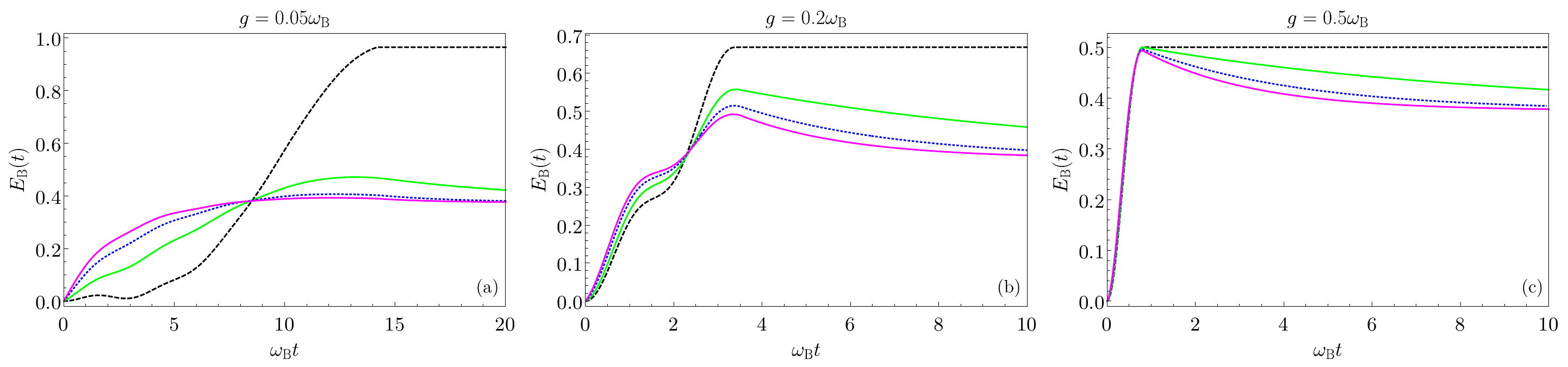} 
   \caption {Behaviour of $E_{\rm B}(t)$ (in units of $\omega_{\rm B}$) as function of $\omega_{\rm B}t$ for a Fock state at $N=8$ and three different values of the coupling: $g=0.05\omega_{\rm B}$ (a), $g=0.2\omega_{\rm B}$ (b) and $g=0.5\omega_{\rm B}$ (c). We consider, at fixed $\alpha_2=0.01$, different values of $\alpha_1$: $\alpha_1=0.03$ (full green curves), $\alpha_1=0.07$ (dotted blue curves) and $\alpha_1=0.1$ (full magenta curves). The dashed black curves are the reference case in absence of dissipation ($\alpha_1=\alpha_2=0$). Other parameters are $\omega_{\rm C}=\omega_{\rm M}=\omega_{\rm B}$, $N_{\rm max} =10 N$, $\beta \omega_{\rm B}=10$ and $\omega_{\rm cut}=500\omega_{\rm B}$. }
   \label{figA1}
\end{figure}

\end{widetext}

In this Appendix we consider the effects of the two thermal baths, when as initial state of the cavity it is considered a Fock state. 
The qualitative behaviour is in general identical to the one obtained in the main text for the coherent state. However, the Fock state is more unstable in presence of dissipation.
This can be seen from Fig.~\ref{figA1}, where the behaviour of $E_{\rm B}(t)$ is reported for the three couplings considered in Sec.~\ref{Dissipation}.

In fact, it is possible to observe that, when the coupling between each part of the system is weak, the energy is suppressed, even for very low dissipation strengths ($\alpha_1=0.03$ and $\alpha_2=0.01$). Increasing the coupling strength allows to get better results compared to the case without dissipation, reaching the best stability in the USC regime at $g=0.5\omega_{\rm B}$ [see panel (c)]. However, we recall that at such high couplings the Fock state has poor performances even without dissipation when one considers the maximum of the transferred energy, i.e. $E^0_{\rm B, max}\sim 0.50\omega_{\rm B}$.

\begin{figure}[h!]
\centering
\includegraphics[scale=0.5]{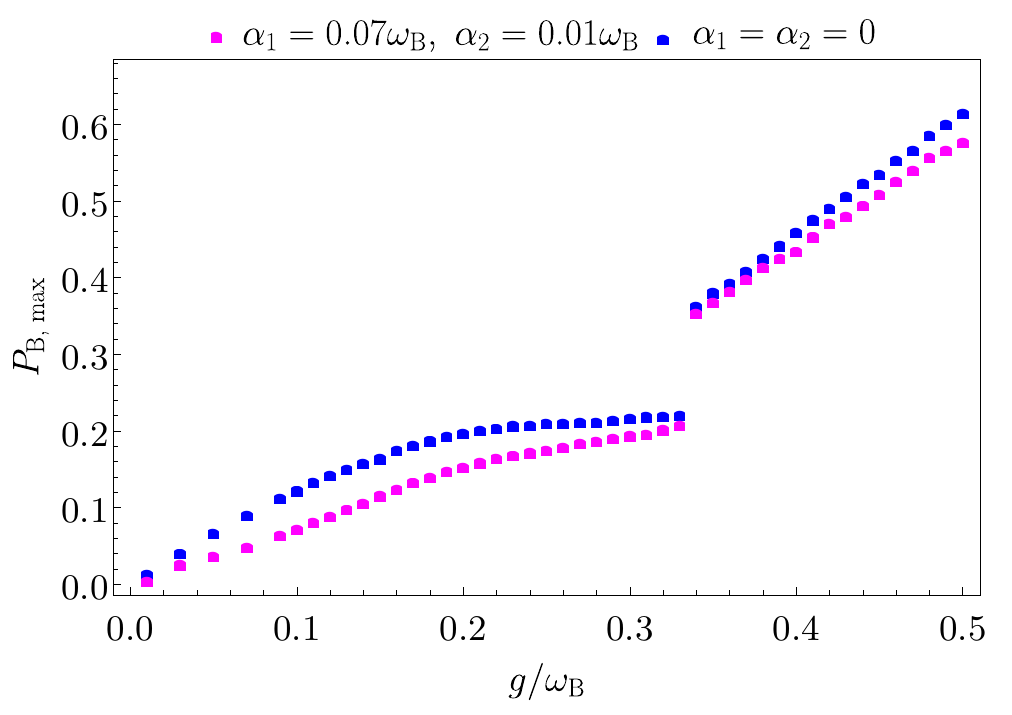} 
   \caption {Behaviour of $P_{\rm B, max}^{\rm d}$ (in units of $\omega_{\rm B}^2$) as function of $g/\omega_{\rm B}$ for a Fock state at $N=8$, for the dissipative case at $\alpha_1=0.07$ and $\alpha_2=0.01$ (magenta square) in comparison with the case with no dissipation at $\alpha_1=\alpha_2=0$ (blue square). Other parameters are $\omega_{\rm C}=\omega_{\rm M}=\omega_{\rm B}$, $N_{\rm max} =10 N$, $\beta \omega_{\rm B}=10$ and $\omega_{\rm cut}=500\omega_{\rm B}$.}
   \label{figA2}
\end{figure}

To conclude the analysis we also consider the dissipative effects of the two thermal baths on the average charging power, focussing on the representative case with $\alpha_1=0.07\omega_{\rm B}$ and $\alpha_2=0.01\omega_{\rm B}$ and compare it to the non dissipative case (see Fig.~\ref{figA2}). Again, the qualitative behaviour is completely analogous to the one obtained for the coherent state. However, being the energy transferred from the quantum charger to the QB lower in the latter case, and having longer transfer times, the average charging power is considerably lower.
Moreover, it is still possible to observe the jump of value at $g^*=0.34\omega_{\rm B}$, meaning that this important feature is not suppressed by dissipation also with Fock state in the considered range of parameters. Finally, we see that the dissipation has a slightly higher impact on the value of $P_{\rm B, max}$. In fact, a relevant discrepancy can be seen in the regimes $0.05 \omega_{\rm B}\lesssim g\lesssim 0.3\omega_{\rm B}$. Then, around the critical value the data almost coincides with the non dissipative case, while for higher couplings the two start to differ again.

This allows us to state that dissipation has more impact on the performances of the Fock state, proving the interest in engineering coherent state for future devices.



\newpage
\begin{thebibliography}{100}
%
\bibitem{Dowling03} J. P. Dowling and G. J. Milburn, Quantum technology: the second quantum revolution, Phil. Trans. R. Soc. Lond. A \textbf{361}, 1655-1674 (2003).
%
\bibitem{Born26} M. Born, Zur Quantenmechanik der Sto\ss vorg\"ange, Z. Physik \textbf{37}, 863-867 (1926). 
%
\bibitem{OBrien09} J. L. O'Brien, A. Furusawa, and J. Vu\v ckovi\'c, Photonic quantum technologies, Nat. Photonics \textbf{3}, 687-695 (2009).
%
\bibitem{Riedel17} M. F. Riedel, D. Binosi, R. Thew, and T. Calarco, The European quantum technologies flagship programme, Quantum Sci. Technol. \textbf{2}, 030501 (2017).
%
\bibitem{Acin18} A. Ac\'in \textit{et al.}, The quantum technologies roadmap: a European community view, New J. Phys. \textbf{20}, 080201 (2018).
%
\bibitem{Zhang19a} Q. Zhang, F. Xu, L. Li, N.-L. Liu, and J.-W. Pan, Quantum information research in China, Quantum Sci. Technol. \textbf{4}, 040503 (2019).
%
\bibitem{Raymer19} M. G. Raymer and C. Monroe, The us national quantum initiative, Quantum Sci. Technol. \textbf{4}, 020504 (2019).
%
\bibitem{Porta20} S. Porta, F. Cavaliere, M. Sassetti, and N. Traverso Ziani, Topological classification of dynamical quantum phase transitions in the xy chain, Sci Rep \textbf{10}, 12766 (2020).
%
\bibitem{Wang20} J. Wang, F. Sciarrino, A. Laing, and M. G. Thompson, Integrated photonic quantum technologies, Nat. Photonics \textbf{14}, 273-284 (2020).
%
\bibitem{Vinjanampathy16} S. Vinjanampathy and J. Anders, Quantum thermodynamics, Contemp. Phys. \textbf{57}, 545 (2016).
%
\bibitem{Campisi17} M. Campisi and J. Goold, Thermodynamics of quantum information scrambling, Phys. Rev. E \textbf{95}, 062127 (2017).
%
\bibitem{Bera19} M. N. Bera, A. Riera, M. Lewenstein, Z. B. Khanian, and A. Winter, Thermodynamics as a consequence of information conservation, Quantum \textbf{3}, 121 (2019).
%
\bibitem{Campisi16} M. Campisi and R. Fazio, Dissipation, correlation and lags in heat engines, J. Phys. A: Math. Theor. \textbf{49}, 345002 (2016).
%
\bibitem{Benenti17} G. Benenti, G. Casati, K. Saito, and R. S. Whitney, Fundamental aspects of steady-state conversion of heat to work at the nanoscale, Phys. Rep. \textbf{694}, 1 (2017).
%
\bibitem{Carrega19} M. Carrega, M. Sassetti, and U. Weiss, Optimal work-to-work conversion of a nonlinear quantum Brownian duet, Phys. Rev. A \textbf{99}, 062111 (2019).
%
\bibitem{Vischi19} F. Vischi, M. Carrega, P. Virtanen, E. Strambini, A. Braggio, and F. Giazotto, Thermodynamic cycles in Josephson junctions, Sci. Rep. \textbf{9}, 3238 (2019).
%
\bibitem{Alicki13} R. Alicki and M. Fannes, Entanglement boost for extractable work from ensembles of quantum batteries, Phys. Rev. E \textbf{87}, 042123 (2013).
%
\bibitem{Campaioli_book} F. Campaioli, F. A. Pollock, and S. Vinjanampathy, \textit{Thermodynamics in the Quantum Regime}, edited by F. Binder, L. A. Correa, C. Gogolin, J. Anders, and G. Adesso (Springer, Berlin, 2018).
%
\bibitem{Campaioli17} F. Campaioli, F. A. Pollock, F. C. Binder, L. Celeri, J. Goold, S. Vinjanampathy,
and K. Modi, Enhancing the Charging Power of Quantum Batteries, Phys. Rev. Lett. \textbf{118}, 150601 (2017).
%
\bibitem{Andolina18} G. M. Andolina, D. Farina, A. Mari, V. Pellegrini, V. Giovannetti, and M. Polini, Charger-Mediated energy transfer in exactly solvable models for quantum batteries, Phys. Rev. B \textbf{98}, 205423 (2018).
%
\bibitem{Farina19} D. Farina, G. M. Andolina, A. Mari, M. Polini, and V. Giovannetti, Charger-mediated energy transfer for quantum batteries: An open-system approach, Phys. Rev. B \textbf{99}, 035421 (2019).
%
\bibitem{Zhang19} Y.-Y. Zhang, T.-R. Yang, L. Fu, and X. Wang, Powerful harmonic charging in a quantum battery, Phys. Rev. E \textbf{99}, 052106 (2019).
%
\bibitem{Chen20} J. Chen, L. Zhan, L. Shao, X. Zhang, Y.-Y. Zhang, and X. Wang, Charging Quantum Batteries with a General Harmonic Driving Field, Ann. Physik \textbf{532}, 1900487 (2020).
%
\bibitem{Crescente20} A. Crescente, M. Carrega, M. Sassetti, and D. Ferraro, Charging and energy fluctuations of a driven quantum battery, New J. Phys. \textbf{22}, 063057 (2020).
%
\bibitem{Carrega20} M. Carrega, A. Crescente, D. Ferraro, and M. Sassetti, Dissipative dynamics of an open quantum battery, New J. Phys. \textbf{22}, 083085 (2020).
%
\bibitem{Le18} T. P. Le, J. Levinsen, K. Modi, M. M. Parish, and F. A. Pollock, Spin-chain model of a many-body quantum battery, Phys. Rev. A \textbf{97}, 022106 (2018).
%
\bibitem{Ferraro18} D. Ferraro, M. Campisi, G. M. Andolina, V. Pellegrini, and M. Polini, High-Power Collective Charging of a Solid-State Quantum Battery, Phys. Rev. Lett. \textbf{120}, 117702 (2018).
%
\bibitem{Andolina19} G. M. Andolina, M. Keck, A. Mari, V. Giovannetti, and M. Polini, Quantum versus classical many-body batteries, Phys. Rev. B \textbf{99}, 205437 (2019).
%
\bibitem{Crescente20b} A. Crescente, M. Carrega, M. Sassetti, and D. Ferraro, Ultrafast charging in a two-photon Dicke quantum battery, Phys. Rev. B \textbf{102}, 245407 (2020).
%
\bibitem{Rossini20} D. Rossini, G. M. Andolina, D. Rosa, M. Carrega, and M. Polini, Quantum Advantage in the Charging Process of Sachdev-Ye-Kitaev Batteries, Phys. Rev. Lett. \textbf{125}, 236402 (2020).
%
\bibitem{Rosa20} D. Rosa, D. Rossini, G. M. Andolina, M. Polini, and M. Carrega, Ultra-stable charging of fast-scrambling SYK quantum batteries, J. High Energ. Phys. \textbf{2020}, 67 (2020).
%
\bibitem{Quach20} J. Q. Quach and W. J. Munro, Using Dark States to Charge and Stabilize Open Quantum Batteries, Phys. Rev. Applied \textbf{14}, 024092 (2020).
%
\bibitem{Santos21} A. C. Santos, Quantum advantage of two-level batteries in the self-discharging process, Phys. Rev. E \textbf{103}, 042118 (2021).
%
\bibitem{Peng21} L. Peng, W.-B. He, S. Chesi, H.-Q. Lin, and X.-W. Guan, Lower and upper bounds of quantum battery power in multiple central spin systems, Phys. Rev. A \textbf{103}, 052220 (2021).
%
\bibitem{Andolina19b} G. M. Andolina, M. Keck, A. Mari, M. Campisi, V. Giovannetti, and M. Polini, Extractable Work, the Role of Correlations, and Asymptotic Freedom in Quantum Batteries, Phys. Rev. Lett. \textbf{122}, 047702 (2019).
%
\bibitem{Dou22} F.-Q. Dou, Y.-J. Wang, and J.-A. Sun, Highly efficient charging and discharging of three-level quantum batteries through shortcuts to adiabaticity, Front. Phys. \textbf{17}, 31503 (2022).
%
\bibitem{Quach22} J. Q. Quach, K. E. Mcghee, L. Ganzer, D. M. Rouse, B. W. Lovett, E. M. Gauger,
J. Keeling, G. Cerullo, D. G. Lidzey, and T. Virgili, Superabsorption in an organic microcavity: Toward a quantum battery, Sci. Adv. \textbf{8}, eabk3160 (2022).
%
\bibitem{Hu22} C.-K. Hu \textit{et al.}, Charging and self-discharging process of a quantum battery in composite environments, Quantum Sci. Technol. \textbf{7}, 045018 (2022).
%
\bibitem{Mailette22} I. Maillette \textit{et al.}, Experimental analysis of energy transfers between a quantum emitter and light fields, arXiv:2202.01109. 
%
\bibitem{Gemme22} G. Gemme, M. Grossi, D. Ferraro, S. Vallecorsa, and M. Sassetti, IBM quantum platforms: a quantum battery perspective, Batteries \textbf{8}, 43 (2022).
%
\bibitem{Gemme23} G. Gemme, M. Grossi, S. Vallecorsa, M. Sassetti, and D. Ferraro, Qutrit quantum battery: comparing different charging protocols, arXiv:2306.14537.
%
\bibitem{Centrone23} F. Centrone, L. Mancino, and M. Paternostro, Charging batteries with quantum squeezing, Phys. Rev. A \textbf{108}, 052213 (2023).
%
\bibitem{Crescente22} A. Crescente, D. Ferraro, M. Carrega, and M. Sassetti, Enhancing coherent energy transfer between quantum devices via a mediator, Phys. Rev. Research \textbf{4}, 033216 (2022).
%
\bibitem{Crescente23} A. Crescente, D. Ferraro, M. Carrega, M. Sassetti, Analytically Solvable Model for Qubit-Mediated Energy Transfer between Quantum Batteries, Entropy \textbf{25}, 758 (2023).
%
\bibitem{Niemczyk10} T. Niemczyk \textit{et al.}, Circuit quantum electrodynamics in the ultrastrong-coupling regime, Nature Phys. \textbf{6}, 772-776 (2010).
%
\bibitem{Yoshihara16} F. Yoshihara, T. Fuse, S. Ashhab, K. Kakuyanagi, S. Saito, and K. Semba, Nature Phys. \textbf{13}, 44-47 (2017).
%
\bibitem{Kockum19} A. F. Kockum, A. Miranowiez, S. De Liberato, S. Savasta, and F. Nori, Ultrastrong coupling between light and matter, Nat. Rev. Phys. \textbf{1}, 19-40 (2019).
%
\bibitem{Beaudoin11} F. Beaudoin, J. M. Gambetta, A. Blais, Dissipation and ultrastrong coupling in circuit QED, Phys. Rev. A \textbf{84}, 043832 (2011).
%
\bibitem{Felicetti14} S. Felicetti, G. Romero, D. Rossini, R. Fazio, and E. Solano, Photon transfer in ultrastrongly coupled three-cavity arrays, Phys. Rev. A \textbf{89}, 013853 (2014).
%
\bibitem{Peng14} J. Peng, Z. Ren, D. Braak, G. Guo, G. Ju X. Zhang, and X. Guo, Solution of the two-qubit quantum Rabi model and its exceptional eigenstates, J. Phys. A: Math. Theor. \textbf{47}, 265303 (2014).
%
\bibitem{Chilingaryan13} S. A. Chilingarayan and B. M. Rodr\'iguez-Lara, The quantum Rabi model for two qubits, J. Phys. A: Math. Theor. \textbf{46}, 335391 (2013).
%
\bibitem{Qu20} P. Qu and Z. Yan, Numerical Calculation of Two-qubit Rabi Model, IOP Conf. Series: Materials Science and Engineering \textbf{994}, 012014 (2020).
%
\bibitem{Defilippis23} G. De Filippis, A. de Candia, G. Di Bello, C.?A. Perroni, L.?M. Cangemi, A. Nocera, M. Sassetti, R. Fazio, and V. Cataudella, Signatures of Dissipation Driven Quantum Phase Transition in Rabi Model, Phys. Rev. Lett. \textbf{130}, 210404 (2023).
%
\bibitem{Schweber67} S. Schweber, On the application of Bargmann Hilbert spaces to dynamical problems, Ann. Phys. \textbf{41}, 205 (1967).
%
\bibitem{Graham84} R. Graham and M. Höhnerbach, Two-state system coupled to a boson mode: Quantum dynamics and classical approximations, Z. Phys. B \textbf{57}, 233 (1984).
%
\bibitem{Schleich} W. P. Schleich, Quantum Optics in Phase Space (Wiley-VCH, Berlin, 2021).
%
\bibitem{Sillanpaa07} M. A. Sillanp\"a\"a, J. I. Park, and R. W. Simmonds, Coherent quantum state storage and transfer between two phase qubits via a resonant cavity, Nature \textbf{449}, 438 (2007).
%
\bibitem{HCollard22} P. Harvey-Collard, J. Dijkema, G. Zheng, A. Sammak, G. Scappucci, and L. M. K. Vandersypen, Coherent Spin-Spin Coupling Mediated by Virtual Microwave Photons, Phys. Rev. X \textbf{12}, 021026 (2022). 
%
\bibitem{Weiss} U. Weiss, \textit{Quantum Dissipative Systems 4th edn} (World Scientific, Singapore, 2012).
%
\bibitem{Caldeira83} A. O. Caldeira and A. Leggett, Path integral approach to quantum Brownian motion, Physica A \textbf{121}, 587 (1983).
%
\bibitem{Leggett87} A. J. Leggett, S. Chakravarty, A. Dorsey, M. P. Fisher, A. Garg, and W. Zwerger, Dynamics of the dissipative two-state system, Rev. Mod. Phys. \textbf{59}, 1 (1987).
%
\bibitem{Ingold} G. L. Ingold, \textit{Path Integrals and Their Application to Dissipative Quantum Systems Coherent Evolution in Noisy Environments} (Springer, Berlin, 2002), pp. 1-53.

%
\bibitem{Rodriguez20} R. H. Rodriguez, F. D. Parmentier, D. Ferraro, P. Roulleau, U. Gennser, A. Cavanna, M. Sassetti, F. Portier, D. Mailly, and P. Roche, Relaxation and revival of quasiparticles injected in an interacting quantum Hall liquid, Nat. Commun. \textbf{11}, 2426 (2020).
%
\bibitem{Dias21} J. Dias, C. W. W\"achtler, V. M. Bastides, K. Nemoto, and W. J. Munro, Reservoir-assisted energy migration through multiple spin domains, Phys. Rev. B \textbf{104}, L140303 (2021).
%
\bibitem{Dias23} J. Dias, C. W. W\"achtler, K. Nemoto, and W. J. Munro, Entanglement generation in never interacting spins via reservoir engineering, arXiv:2306.07507.
%
\bibitem{Lindblad75} G. Lindblad, Completely positive maps and entropy inequalities, Commun. Math. Phys. \textbf{40}, 147-151 (1975).
%
\bibitem{Lindblad76} G. Lindblad, On the generators of quantum dynamical semigroups, Commun. Math. Phys. \textbf{48}, 119-130 (1976).
%
\bibitem{Devoret13} M. H. Devoret and R. J. Schoelkopf, Coupling superconducting qubits via cavity bus, Science \textbf{339}, 1169 (2013).
%
\bibitem{Wendin17} G. Wendin, Quantum information processing with superconducting circuits: A review, Rep. Prog. Phys. \textbf{80}, 106001 (2017).
%
\bibitem{Sete21} E. A. Sete, A. Q. Chen, R. Manenti, S. Kulshreshtha, and S. Poletto, Floating Tunable Coupler for Scalable Quantum Computing Architectures, Phys. Rev. Applied \textbf{15}, 064063 (2021).
%
\bibitem{Campbell23} D. L. Campbell, A. Kamal, L. Ranzani, M. Senatore, and M. D. LaHaye, Modular Tunable Coupler for Superconducting Circuits, Phys. Rev. Applied \textbf{19}, 064043 (2023).
%
\bibitem{Heunisch23} L. Heunisch, C. Eichler, and M. J. Hartmann, Tunable coupler to fully decouple superconducting qubits, arXiv:2306.17007.
%
\bibitem{Sassetti90} M. Sassetti and U. Weiss, Universality in the dissipative two-state system, Phys. Rev. Lett. \textbf{65}, 2262 (1990).
%
\bibitem{Grifoni96} M. Grifoni, M. Sassetti, and U. Weiss, Exact master equations for driven dissipative tight-binding models, Phys. Rev. E \textbf{53}, R2033(R) (1996).
%
\bibitem{Delmonte21} A. Delmonte, A. Crescente, M. Carrega, D. Ferraro, and M. Sassetti, Characterization of a two-photon quantum battery: Initial conditions, stability and work extraction, Entropy \textbf{23}, 612 (2021).
%
\bibitem{Maring17} N. Maring, P. Farrera, K. Kutluer, M. Mazzera, G. Heinze, and H. de Riedmatten, Photonic quantum state transfer between a cold atomic gas and a crystal, Nature \textbf{551}, 485-488 (2017).
%
\bibitem{Kurpiers18} P. Kurpiers \textit{et al.}, Deterministic quantum state transfer and remote entanglement using microwave photons, Nature \textbf{558}, 264-267 (2018).
%
\bibitem{Petruccione}  H.-P. Breuer and F. Petruccione, \textit{The Theory of Open Quantum Systems}, (Oxford University Press, Oxford, 2002).
%
\bibitem{Makhlin03} Y. Makhlin, G. Sch\'on, and A. Shnirman, Dissipative effects in Josephson qubits, Chem. Phys. \textbf{296}, 315-324 (2003).
%
\bibitem{Johansson13} J. R. Johansson, P. D. Nation, and F. Nori, QuTiP 2: A Python framework for the dynamics of open quantum systems, Comput. Phys. Commun. \textbf{184}, 1234 (2013).
%
\bibitem{note1} Notice that the terms $(a+a^\dagger)\sigma_x$ in the Hamiltonian in Eq.~(\ref{Htot}) can be rewritten as $(a+a^\dagger)(\sigma_++\sigma_-)=(a\sigma_++a\sigma_-+a^\dagger\sigma_++a^\dagger\sigma_+)$. Here, the so-called counter-rotating terms $a\sigma_-$ and $a^\dagger\sigma_+$ does not conserve the number of excitations, and are the ones that can be neglected in the rotating-wave approximation. 
%
\bibitem{Stramacchia19} M. Stramacchia , A. Ridolfo, G. Benenti, E. Paladino, F. M. D. Pellegrino, D. Maccarrone, and G. Falci, Speedup of Adiabatic Multiqubit State-Transfer by Ultrastrong Coupling of Matter and Radiation, Proceedings \textbf{12}, 35 (2019).
%
\bibitem{Kirton19} P. Kirton, M. M. Roses, J. Keeling, and E. G. Dalla Torre, Introduction to the Dicke Model: From Equilibrium to Nonequilibrium, and Vice Versa, Adv. Quantum Technol. \textbf 2, 1800043 (2019).
%
\bibitem{Scarlino19} P. Scarlino \textit{et al.}, Coherent microwave-photon-mediated coupling between a semiconductor and a superconducting qubit, Nat. Commun. \textbf{10}, 3011 (2019).









\end{thebibliography}
\end{document}